\documentclass[showpacs,prb,twocolumn]{revtex4}

\usepackage{graphicx,float,amsmath}
\begin{document}

\title{Fowler-Nordheim-like local injection of photoelectrons from a silicon
tip.}
\author{A.C.H. Rowe}
\affiliation{Laboratoire de physique de la mati\`ere condens\'ee, \'Ecole Polytechnique,
91128 Palaiseau Cedex, France}
\author{D. Paget}
\email{daniel.paget@polytechnique.fr}
\affiliation{Laboratoire de physique de la mati\`ere condens\'ee, \'Ecole Polytechnique,
91128 Palaiseau Cedex, France}


\begin{abstract}
Tunneling between a photo-excited p-type silicon tip and a gold surface is studied as a function of tip bias, tip/sample distance and light intensity. In order to extend the range of application of future spin injection experiments, the measurements are carried out under nitrogen gas at room temperature. It is found that while tunneling of valence band electrons is described by a standard process between the semiconductor valence band and the metal, the tunneling of photoelectrons obeys a Fowler-Nordheim-like process \textit{directly} from the conduction band. In the latter case, the bias dependence of the photocurrent as a function of distance is in agreement with theoretical predictions which include image charge effects. Quantitative analysis of the bias dependence of the dark and photocurrent spectra gives reasonable values for the distance, and for the tip and metal work functions. For small distances image charge effects induce a vanishing of the barrier and the bias dependence of the photocurrent is exponential. In common with many works on field emission, fluctuations in the tunneling currents are observed. These are mainly attributed to changes in the prefactor for the tunneling photocurrent, which we suggest is caused by an electric-field-induced modification of the thickness of the natural oxide layer covering the tip apex. \end{abstract}

\maketitle

\section{Introduction}

Although the photo-sensitivity of tunnel processes between a metal and a semiconductor has been widely considered \cite{grafstrom02,glembocki92,hamers90}, the majority of these investigations concern the effect of surface photovoltage on tunneling between a metallic tip and a planar semiconducting surface. On the other hand, the mechanisms governing injection of photoelectrons from a semiconducting tip under light excitation into a planar metallic surface are less well studied despite the fact that a proper understanding is important both from a fundamental viewpoint, and for a variety of potential applications.

Photoexcited semiconducting tips can in principle be used as local spin injectors since circularly polarized light produces a spin polarized electron population via optical pumping\cite{lampel68}. The mean spin of the tunnel injected electrons can then be controlled via a change in the light helicity. There exists numerous potential applications for a spin injector of this type, such as spin-polarized scanning tunneling microscopy (SPSTM)\cite{pierce88}, quantum computing \cite{divincenzo99} and spintronics \cite{datta90}, none of which has been convincingly demonstrated. While it is known that a change in pump light helicity modifies the tunnel junction resistance between a ferromagnetic metallic tip and a photoexcited GaAs surface via the spin-dependence of the tunneling process\cite{alvarado95}, spurious optical effects in photoexcited semiconductor tips have been blamed for the apparent observation of non-zero spin polarizations, even on non-magnetic surfaces \cite{jansen99,nabhan99}. In order to better understand \textit{non-polarized} photo-assisted tunneling from a semiconductor tip into a metallic surface a number of groups\cite{bode03} commenced by studies of the tip bias, distance and light intensity dependence of the injected current. Perhaps the most detailed investigations, both theoretical and experimental, were performed by the Nijmegen group \cite{prins96}. For relatively small tip bias up to about 0.5 V, tunneling current spectra were interpreted using a model that considered both the characteristics of the space charge layer formed at the tip surface, and of the tunnel barrier itself. While this model can account for a number of observed phenomena, its extension to include spin polarization effects has met with less success \cite{jansen99,jansen98}.

A second related area of interest is the emission of electrons from high aspect ratio objects such as semiconducting\cite{miyamoto03,gunther03,forbes99} or carbon nanotube tips\cite{bonard02}, as well as the influence of photo-excitation on the emitted current \cite{jensen00}. While the electric fields are similar to those used in tunneling experiments, the tip/sample (or rather the cathode/anode) distances are typically much larger (up to several $\mu$m). Under these conditions, the emitted current is usually described by a Fowler-Nordheim (FN) process \cite{fowler28} from surface states to vacuum states. Unless a complex sample (anode) structure is used\cite{filipe98}, this process is spin independent since the vacuum states are themselves unpolarized. The bias dependence of the FN current is of the form \cite{fowler28,schwettman74}
\begin{equation}
I=A(V)/(\gamma V)^{2}\times \exp (-B(V)/\gamma V), \label{Fowler}
\end{equation}
where $A(V)$ and $B(V)$ are slowly-varying functions of bias. The quantity $\gamma$, larger than unity for sharp tips, describes the geometrical enhancement of the electric field\cite{schwettman74}.

\begin{figure}[tbp]
\includegraphics[clip,width=5 cm] {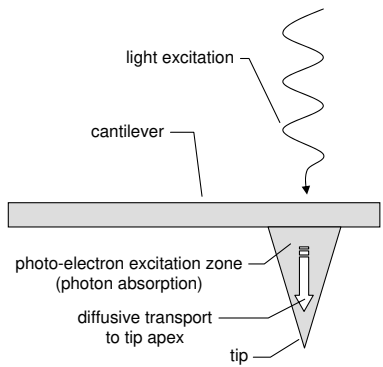}
\caption{Injection geometry: a semiconducting tip, placed at the end of an
AFM-like cantilever, is excited by light impinging on its rear planar face.}
\label{figure1}
\end{figure}

In the present paper we investigate the mechanisms which govern photo-electron injection from a semiconducting tip. The tip bias value, as large as -3 V, is sufficient to induce an increase in the injected current by FN-tunneling, but not large enough to enable observation of current oscillations related to quantized states in the tunneling gap\cite{becker85}. Here, we are not interested in spin-polarized injection so silicon tips and gold surfaces are used. With respect to a number of previous works on tunnel injection of photocarriers \cite{jansen98, jansen99, prins96, suzuki99, nabhan99}, the present situation is simpler since: i) the light excitation is incident on the \textit{rear planar surface} of the tip, which itself is situated at the end of a very stiff cantilever (see Fig. \ref{figure1}). Consequently, the photoelectrons diffuse from the rear of the tip (where they are created) to the tip apex, and the number of photoelectrons reaching the tip apex is independent of tip bias. This is at variance with studies where the light excitation is incident on the front face of the tip apex\cite{jansen98,prins96}, and the injected photocurrent depends on tip bias via the bias dependent depletion layer width\cite{gartner59}. The experiment reported here thus resembles an atomic force microscope (AFM), with the exception that tip/sample distance is controlled using the tunnel current \cite{footnote1}, and ii) the use of p$^{+}$ doped tips guaranties that the conductivity of the tip is high and that most of the applied bias is dropped across the tunnel barrier itself. The experimental results can then be understood using a relatively simple theory, iii) the tip surface barrier ($V_{b}$) can be neglected for $V > V_{b}$, and iv) the experiments are performed in an inert gas atmosphere. Although this has intrinsic drawbacks compared with ultra-high vacuum conditions, in particular related to possible tip pollution and current instabilities caused by changes in the native oxide layer thickness, the choice is deliberately made in order to simplify and extend the scope of application of future SPSTM studies\cite{paget05}.

The dependence of the tunneling dark current and photocurrent as a function of bias, tip/sample distance and excitation light intensity is investigated. It is found that i) the dark current obeys a standard tunneling process between the valence band of the semiconductor and the metallic local density-of-states, and ii) the tunneling photocurrent occurs directly from the conduction band of the semiconductor to the metallic states through a FN-like process and not, as found by the Nijmegen group for GaAs tips \cite{prins96}, via a standard tunneling process from midgap states at the semiconductor surface. The implications of these findings for spin-polarized tunneling are discussed.

\section{Theoretical Considerations}

\begin{figure}[tbp]
\includegraphics[clip,width=8 cm] {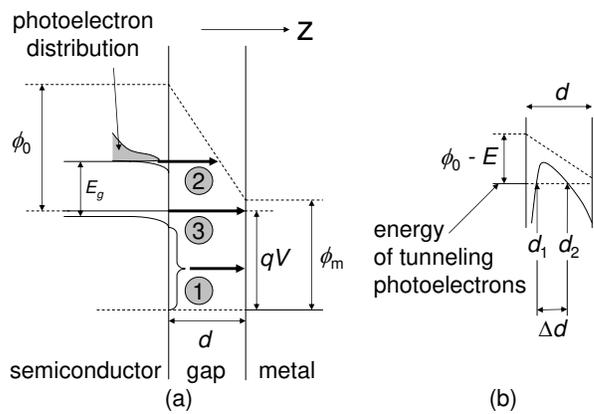}
\caption{(a) Schema of the band structure of the semiconductor (left) and of the metal (right) for a tip bias $V$. Arrow 1 represents the dark current arising from a tunneling process between the semiconductor valence band and the metal. Arrow 3 represents the current arising from a tunneling process between localized mid-gap semiconductor states and the metal, as proposed in the Nijmegen model\cite{prins96}. Arrow 2 shows the observed FN-like tunneling process of photoelectrons from the semiconductor conduction band. (b) Tunnel barrier shape with (solid line) and without (dotted line) image charge effects.}
\label{figure2}
\end{figure}

The following section describes tunnel injection between a negatively biased, p$^{+}$ doped photoexcited semiconducting tip and a metallic surface. Shown in Fig. \ref{figure2}a is a schema of the band structure of the tip and metallic surface at applied tip bias, $V$. Light excitation from the rear of the tip results in the formation of a steady state population of photoelectrons in the conduction band. Since the tip is highly conductive, the semiconductor band structure is independent of bias and is simply shifted with respect to the metallic one by $qV$, where $q<0$ is the electronic charge.

Shown in Fig. \ref{figure2}b is a schematic of the shape of the tunneling barrier. Because of image charge effects\cite{simmons63}, the tunneling barrier at the energy of the tunneling electrons extends between $d_{1}$ and $d_{2}$, and its width $\Delta d=d_{2}-d_{1}$ is smaller than the physical tip/sample distance, $d$.

Because the applied tip bias is negative, the dark current and the current of minority photocarriers are described by a tunneling of electrons \textit{from} the semiconductor \textit{to} the metal. The incremental tunnel current from an occupied energy level $E$ in the tip to an empty level in the metal at the same energy is given by\cite{simmons63,chen} 
\begin{equation}
\delta I_{t}=KS\rho _{s}\rho _{m}\exp (-\kappa \Delta d), \label{infcurr}
\end{equation}
where $\rho _{s}$ and $\rho _{m}$ are the density of states at $E$ of the semiconductor and of the metal respectively, and $S$ is the tip surface area. The quantity $K$ is proportional to the tunneling matrix element, and depends on the electronic orbitals of tip and surface via which tunneling occurs. The inverse distance for the exponential decay of the tunneling current is given by\cite{simmons63} 
\begin{equation}
\kappa =2\frac{\sqrt{2m}}{\hbar }\sqrt{\overline{\phi }-E}, \label{d0}
\end{equation}
where $m$ is the mass of the electron in vacuum. Here, the zero of energy is taken at the Fermi level of the semiconductor. In the work of Simmons \cite{simmons63}, $E$ is the sum of the potential energy of the tunneling electrons and of their kinetic energy oriented perpendicular to the surface $E_{z}$. In the present case, since tunneling from a tip is essentially a unidirectional process\cite{footnote2}, the tunneling electrons have a velocity along the tip axis and $E_{z}=E$. Eq. \ref{d0} expresses the highly simplifying result that tunneling across a barrier of complex spatial shape mostly depends on the spatial average of the barrier, defined as 
\begin{equation}
\overline{\phi }=\frac{1}{\Delta d}\int_{d_{1}}^{d_{2}}\phi (z)dz.
\label{phibar}
\end{equation}

\subsection{Tunneling photocurrent}

In general the tunneling current obtained from Eq. \ref{infcurr}, Eq. \ref{d0} and Eq. \ref{phibar} is given by the expression
\begin{equation}
I_{ph}=A_{ph}\exp (-\kappa _{ph}\ d). \label{Iphstandard}
\end{equation}
The prefactor $A_{ph}$ is found by replacing $\rho_{s}$ by the concentration $N_{e}$ of photoelectrons (a quantity proportional to the excitation light power):
\begin{equation}
A_{ph}=K_{ph}SN_{e}\rho _{m}. \label{prefactor}
\end{equation}
For a standard tunneling current in the absence of image charge effects one has $\overline{\phi } =\phi _{0}-qV^{\ast}/2$. Since the photoexcitation is only slightly above gap $E$ is the semiconductor bandgap energy $E_{g}$ and $\kappa _{ph}$ is given by 
\begin{equation} \kappa _{ph}=2 \sqrt{\frac{2m}{\hbar}(\chi_{0}-qV^{\ast}/2)},
\label{d0photo1}
\end{equation}
where $\chi_{0}= \phi _{0}-E_{g}$ is the affinity of the tip surface. The effective tip bias $V^{\ast}$ is given by
\begin{equation} V^{\ast}=V+V_{0}=V+\frac{1}{q} \left(\phi_{0}-\phi_{m}\right) \label{Vstar}
\end{equation}
since there is an electric field at $V=0$ between tip and surface if $\phi _{m}$ is different from $\phi _{0}$ ($\phi _{0}-\phi _{m} \approx -0.6$ eV for clean surfaces of silicon and gold).

At larger tip bias, tunneling from the semiconductor to the metal can occur via vacuum states (see  Fig. \ref{figure2}a, arrow 2). This process resembles FN emission with three differences with respect to Eq. \ref{Fowler} and will therefore be denominated \textit{FN-like}. Firstly, in cases where the tip/sample distance is small compared with the radius of curvature of the tip, no field enhancement occurs at the semiconducting tip (i.e. $\gamma$=1 in Eq. \ref{Fowler} as explained in Appendix 1). A second difference with the field emission process is the replacement of $V$ by $V^{\ast}$ given by Eq. \ref{Vstar} (in field emission studies $\phi_{0}-\phi _{m}$ is negligible compared to the applied tip (cathode) bias and can be ignored). Finally, since the tunneling photoelectrons are distributed over a narrow energy range at the bottom of the conduction band, it is not necessary to perform an integration over energy. Unlike Eq. \ref{Fowler}, the prefactor is independent of tip bias so that Eq. \ref{Iphstandard} is still valid, albeit with a modified value of $\kappa_{ph}$.

\subsubsection{FN-like tunneling without image charge effects}

In order to observe FN-like tunneling, $qV^{\ast}$ must be larger than $\phi_{0} - E_{g}$, so that the threshold bias $V_{th}$ is
\begin{equation}
qV_{th}= \phi_{m}-E_{g}. \label{seuilloin}
\end{equation}
The spatial average $\overline{\phi}$ of the barrier potential\cite{footnote3} is then equal to $\left(\phi_{0} +E_{g}\right)/2$. The tunneling photocurrent is obtained from a calculation of the quantum tunneling probability at $E_{g}$, and is given by Eq. \ref{Iphstandard} with 
\begin{equation}
\kappa_{ph}= 2\frac{\sqrt{m}}{\hbar}\frac{\chi_{0}^{3/2}}{qV^{\ast}}. \label{Iphfar}
\end{equation}
If $V_{0}$ is negligible with respect to $V$, Eq. \ref{Iphfar} indicates that a plot of $I_{ph}$ in logarithmic units as a function of $1/V$ should be linear. The slope of this line should decrease when the $d$ is decreased. For a bias lower than $V_{th}$, the photocurrent has a smaller value obtained from Eq. \ref{Iphstandard} and Eq. \ref{d0photo1}.

\subsubsection{FN-like tunneling with image charge effects}

Inclusion of image charge effects requires a modification of Eq. \ref{Iphfar}. It is not a bad approximation to apply the treatment of Simmons for tunneling between two metals\cite{simmons63}, since with respect to the static dielectric constant of vacuum ($\epsilon$ = 1), that of silicon ($\epsilon $ = 13) can be approximated by that of a metal ($\epsilon \rightarrow \infty$). The image charge potential is given by\cite{simmons63} 
\begin{equation}
\phi (z) =\phi_{0} - \frac{qV^{\ast}z}{d} - 1.15\frac{\lambda d^{2}}{z(d-z)}, \label{phiparab}
\end{equation}
where $\lambda$ describes the magnitude of the image charge effects: 
\begin{equation}
\lambda =\frac{q^{2}\ln (2)}{8\pi \epsilon \epsilon_{0}d}. \label{lambda}
\end{equation}
Here $\epsilon_{0}$ is the permittivity of free space. For $qV^{\ast}$ sufficiently large\cite{footnote13} and $d_{1} \ll d$, one finds that the inverse distance for tunneling is now given by 
\begin{equation}
\kappa_{ph}= 2\frac{\sqrt{m}}{\hbar}\frac{\chi
_{0}^{3/2}(1-5.6\lambda/\chi_{0})}{qV^{\ast}}\sqrt{1+5.6\lambda
/\chi_{0}-aqV^{\ast}/\chi_{0}}, \label{Iphmid}
\end{equation}
where
\begin{equation}
a=\frac{\lambda }{\chi_{0}} \left[1.2-\frac{2.3}{1-5.6\lambda
/\chi_{0}}\ln \left(\frac{d_{2}d}{d_{1}(d-d_{2})}\right) \right],
\label{phibarsimp}
\end{equation}
is weakly dependent on $V^{\ast}$ via the logarithmic term. An approximate\cite{footnote13} but physically meaningful expression for the threshold bias $V_{th}$ for FN-like tunneling is found by setting $d_{2}=d$ which gives
\begin{equation}
qV_{th}= \phi_{m}-E_{g}-5.6\lambda, \label{seuil}
\end{equation}
which decreases with decreasing $d$ unlike the value given by Eq. \ref{seuilloin}. As expected, if $\lambda \ll \chi_{0}$, Eq. \ref{Iphmid} reduces to Eq. \ref{Iphfar} and Eq. \ref{seuil} reduces to Eq. \ref{seuilloin}.

The inclusion of image charge effects modifies the photocurrent in two ways: i) \textit{at small bias}: because of the dependence of $\kappa_{ph}$ on $d$ in Eq. \ref{Iphmid}, the slope of $\log(I_{ph})$ as a function of $1/V$ is no longer proportional to $d$ unless this distance is large. The effective tip affinity $\chi_{0}^{\ast}$ is now given by $\chi_{0}^{\ast 3}= \left(\chi_{0}+5.6\lambda \right) \left(\chi_{0}-5.6\lambda \right)^{2}$, and ii) if $aqV^{\ast} \approx \chi_{0} + 5.6\lambda$ the lowering of the barrier due to image charges causes an excess photocurrent.

\subsection{Tunneling dark current}

As represented in Fig. \ref{figure2}, the dark current is presumed to occur from tunneling processes between occupied states in the semiconductor valence band and empty states in the metal at the same energy. At valence band energies, image charge effects\cite{footnote14} and FN-like tunneling can be neglected. An energy-independent density of states for the metal is assumed\cite{papaconstantopoulos86}, and the usual form of $\rho_{s}$ for a bulk semiconductor is used, $\rho_{s}(E)=\frac{1}{2\pi^{2}}[\frac{2m^{\ast}}{\hbar^{2}}]^{3/2}\sqrt{-E}$, where $\hbar$ is Planck's constant, $m^{\ast}$ is the effective electronic mass, and $E<0$ is the energy of the tunneling electrons in the valence band. The value of the dark current is obtained by integrating Eq. \ref{infcurr} between the Fermi level of the semiconductor and that of the metal, 
\begin{equation}
I_{dark}=A_{dark}\int_{-qV}^{0}\sqrt{-E}\exp (-\kappa_{dark}d)dE
\label{Idark}
\end{equation}
where $\kappa_{dark}$ and $A_{dark}$ are given respectively by
\begin{equation}
\kappa _{dark}=2\sqrt{\frac{2m}{\hbar}(\phi_{0}-E-qV^{\ast}/2)}, 
\label{d0dark}
\end{equation}
and
\begin{equation}
A_{dark}=K_{dark}\frac{S\rho_{m}}{2\pi ^{2}}[\frac{2m^{\ast}}{\hbar^{2}}%
]^{3/2}. \label{prefactdark}
\end{equation}

The dark current is the sum of several exponential contributions and generally does not have a simple exponential dependence as a function of distance.

\section{Experimental Details}

\begin{figure}[tbp]
\includegraphics[clip,width=7 cm] {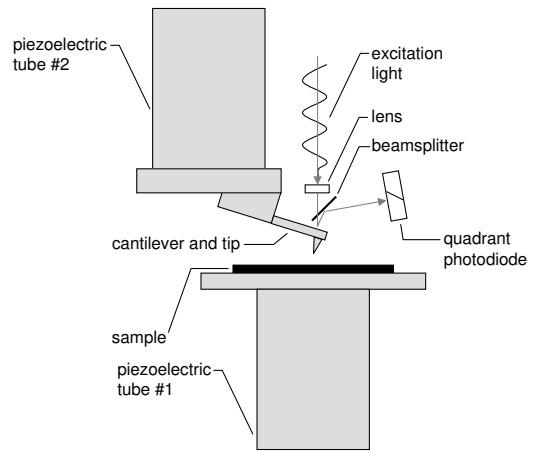}
\caption{Schema of the atomic force microscope-like experimental setup. 
The sample is fixed on a piezoelectric tube, enabling scans in the plane of the sample. The cantilever holding the tip is fixed on a second piezoelectric tube enabling motions in the vertical direction. }
\label{figure3}
\end{figure}

Experiments were conducted in a tailor-made system enabling STM as well as AFM investigations at room temperature, and consistent with the geometry of photoelectron injection described in Fig. \ref{figure1}. This system, shown in Fig. \ref{figure3}, operates in an electromagnetically shielded inert gas environment. Worth mentioning are the following particularities: i) the excitation laser, (wavelength 780 nm, power 10 mW) is focused on the back planar face of the cantilever, yielding a spot of diameter 50 $\mu $m opposite the tip position\cite{footnote15}, ii) a beamsplitter situated between the laser and the cantilever sends the light reflected from the cantilever surface onto a quadrant photodiode, thus enabling simultaneous atomic force, dark current and photocurrent measurements, iii) two piezoelectric tubes are used. Scans in the plane of the sample for imaging purposes are performed using the four-electrode PZT tube on which the sample is mounted, while the tip/sample distance is changed by a second two-electrode PZT tube on which the block holding the cantilever and tip is mounted. This ensures that the light spot is stationary with respect to the location of the tip, iv) for current measurements the sample is grounded and the bias is applied to the tip. The tip/sample current is monitored using a high gain/low noise amplification circuit placed as close as possible to the tip. After installing the sample and cantilever in an air ambient, a metallic grounded hood is placed over the experiment and dry nitrogen is blown for 20 minutes prior to, and during, the measurements.

Commercially available tapping-mode silicon cantilevers\cite{veeco} were used in all experiments. The use of relatively stiff cantilevers (of nominal force constant 16 N/m) minimizes the effect of atomic forces on tip/sample distance, and thereby improves the control of this distance using the tunnel current in a feedback loop. Silicon tips were situated at the free end of the cantilever and ranged in height between 5-7 $\mu$m, and were p-type doped in the range $10^{20}$ cm$^{-3}$. The samples consist of polycrystalline gold films, of thickness approximately 60 monolayers, deposited on (111) silicon substrates by electrochemical methods \cite{prodhomme06}.

Care was taken in order to minimize fluctuations in the tunnel current due to (photo)electro-chemical reactions which can induce changes in the tip characteristics. Anodic tip potentials (forward bias for p-type tips) were avoided since they are known to rapidly generate a thick layer of oxide which prevents the observation of tunnel currents. Several tip passivation treatments were tried in order to render the tip surface inert, including removal of the native oxide with hydrofluoric acid. However, the native oxide layer yielded the most inert surface, and the results reported here were all obtained with naturally oxidized tips. As will be seen, the presence of residual water can still induce electrochemical changes in the tip surface which produce instabilities in the tunneling current.

\begin{figure}[tbp]
\includegraphics[clip,width=8 cm] {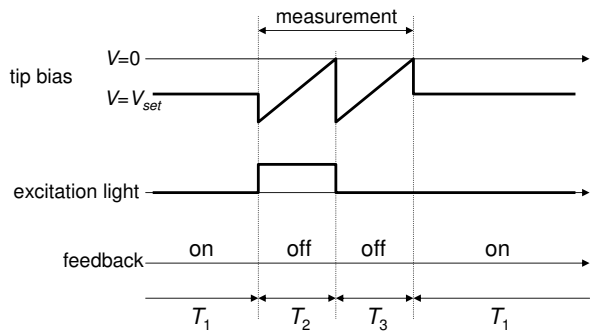}
\caption{Measurement sequence (see text).}
\label{figure4}
\end{figure}

The measurement sequence is shown schematically in Fig. \ref{figure4}. During $T_{1} \approx$ 0.5 s (i.e. much longer than the time constant of the feedback loop) the pump laser is switched off and no data acquisition takes place. The tip/sample distance is stabilized at $V_{set}=-2$ V by imposing a dark tunneling current $I_{dark}=I_{set}$ using a standard feedback loop. Spectroscopic measurements of the tunneling current as a function of tip bias are then performed in a rapid sequence of two acquisitions over periods $T_{2}$ and $ T_{3}$, each of which lasts 25 ms. Since these scans induce a change in current, the feedback loop is opened for $T_{2}$ + $T_{3}=50$ ms, a period significantly shorter than time constants over which drift in $d$ occurs. During $T_{2}$ the laser is switched on, while during $T_{3}$ it is switched off. The tunneling dark current spectrum is that obtained during $T_{3}$, while the tunneling photocurrent spectrum is that obtained by taking the difference of the spectra measured during $T_{2}$ and $T_{3}$. Most of the results presented in this paper were obtained after only one such acquisition (i.e. without averaging over several spectra).

Using this procedure, the dark current, the photocurrent, and the atomic force between tip and surface (measured using the quadrant photodiode), were simultaneously monitored for increasing tunneling dark currents during $T_{1}$, corresponding to decreasing tip/sample distances. In order to check reproducibility and possible current instabilities, each scan was performed ten times in identical conditions.

\section{Results}

\begin{figure}[tbp]
\includegraphics[clip,width=7 cm] {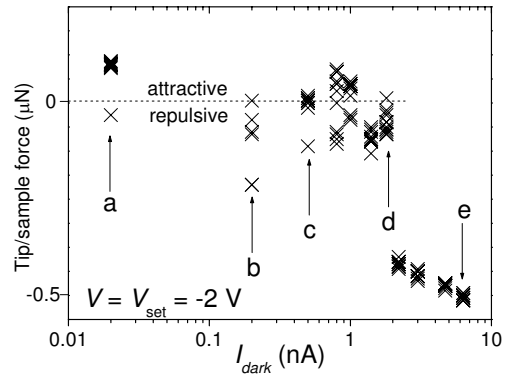}
\caption{Dependence of the atomic force between tip and sample as a function of the dark current value imposed during $T_{1}$. For each value of $I_{dark}$ ten successive measurements were obtained, each of which is shown in the figure. The arrows, a-e correspond to the curves shown in Fig. \ref{figure8}. The abrupt change near $I_{dark}$ = 2 nA is due to mechanical tip/sample contact.}
\label{figure5}
\end{figure}

Fig. \ref{figure5} shows the variation of the atomic force between the tip (shown in the inset of Fig. \ref{figure6}) and the sample, as a function of the tunneling dark current set during time $T_{1}$. Shown by arrows in the figure are the selected dark current values for which tunneling spectra will be presented below. The atomic force stays approximately equal to zero up to a dark current of the order of 2 nA, above which it abruptly becomes repulsive because of mechanical contact with the surface. This abruptness as well as the fact that a tunneling current is observed \textit{before} the onset of a non-zero atomic force are due to the relatively large stiffness of the cantilever. Points a, b, and c correspond to the out of contact regime, while point d illustrates the near-contact regime. At point e, there is no doubt that gold indentation occurs since this has been observed\cite{burnham89} for forces as small as several nN.

\subsection{Instabilities}

\begin{figure}[tbp]
\includegraphics[clip,width=7.5 cm] {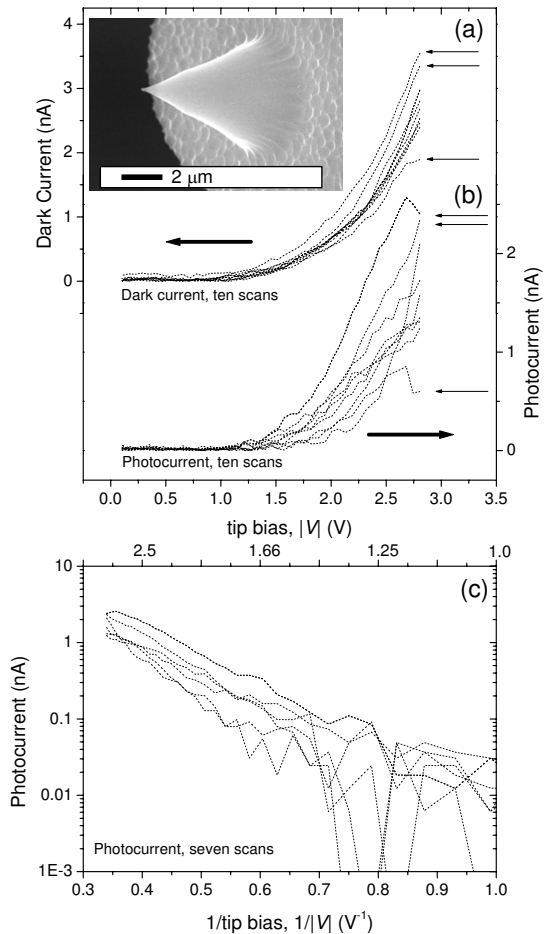}
\caption{Ten successive measurements of (a) dark current and (b) photocurrent as a function of tip bias for fixed tip/sample distance. In (a), three curves (designated by arrows) strongly differ from the other curves at $V_{set}$ and were not considered in the analysis. (c) The linear dependence of the photocurrent of the remaining seven curves on logarithmic plot of the photocurrent versus $1/V$ reveals a FN-like tunnel process. Shown (inset) is a scanning electron microscope image of the sharp tip.}
\label{figure6}
\end{figure}

Shown in Fig. \ref{figure6}a are ten dark current spectra at point c in Fig. \ref{figure5}. The corresponding tunneling photocurrent spectra are shown in Fig. \ref{figure6}b. Both currents strongly increase with bias and no surface photovoltage is observed (no photocurrent is detectable below 1.15 V).

The differences in the curves reveal instabilities in the tunneling process. These instabilities are larger for the tunnel photocurrent, which is consistent with the fact that immediately before the measurement, the feedback imposes a constant dark current $I_{dark}$ at $V = V_{set}$. They manifest themselves as abrupt temporal changes between a relatively small number of well-defined values which can differ by up to a factor of three. Such behaviour has been reported in works on field emission\cite{miyamoto03}, on transport in metal-oxide-silicon structures\cite{farmer87,farmer88}, and in tunneling \cite{rogers85}. Since the delay between the instabilities is of the order of a fraction of a second (i.e. much larger than $T_{2}+T_{3}$), no apparent jump is seen within a given curve. 

Of the ten dark current spectra in Fig. \ref{figure6}a, seven curves approximately coincide. The other three curves are discarded since the current value at $V_{set}$= -2 V differs from the setpoint value. As seen from Eqs. \ref{Iphstandard}-\ref{Iphfar}, Eq. \ref{Idark}, and Eq. \ref{prefactdark}, instabilities can originate from fluctuations in $d$, $(\phi_{0})$ or $(\phi_{m})$, or in the prefactors $A_{ph}$ and $A_{dark}$. Instabilities in the prefactors can be caused by fluctuations of the tip surface $S$ or of the tunneling probabilities $K_{ph}$ or $K_{dark}$ (it is assumed that $N_{e}$ and $\rho_{m}$ are stable).

\begin{figure}[tbp]
\includegraphics[clip,width=8 cm] {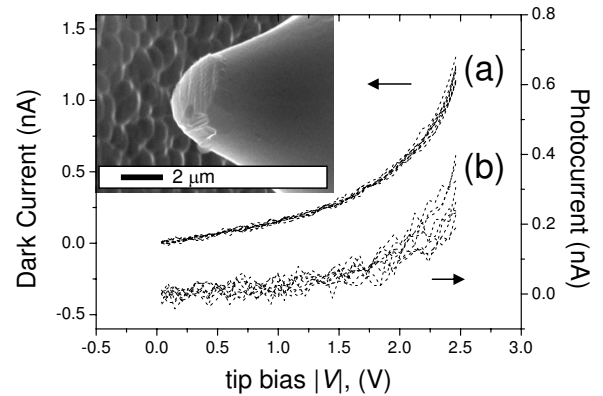}
\caption{Blunt tip: Ten successive measurements of (a) dark current and (b) photocurrent as a function of tip bias for fixed, large tip/sample distance. Shown (inset) is a scanning electron microscope image of the tip.}
\label{figure7}
\end{figure}

Instabilities in $d$ can be ruled out based on results obtained with a blunt tip made by mechanically removing the tip apex. The tip is almost ideally flat, with an end diameter of $\approx$ 2 $\mu$m (see inset of Fig. \ref{figure7}). For a blunt tip, tunnel processes are averaged out over a large area and should therefore be relatively insensitive to local changes in geometry and tip/sample surface composition. The spectra shown in Fig. \ref{figure7} were obtained by fixing the tunneling dark current during $T_{1}$ at the same value used for the sharp tip. The dark current spectra are essentially indistinguishable, while differences between photocurrent spectra are comparable to the noise. The same qualitative results are also obtained for all tip/sample distances. This proves that any tip/sample distance noise or drift is effectively compensated by the feedback loop, and therefore that fluctuations in $d$ are not the root cause of current instabilities measured with the sharp tip.

Fig. \ref{figure6}c shows $I_{ph}$ scans plotted in logarithmic units as a function of $1/V$ for the sharp tip. While $I_{ph}$ varies by up to a factor of 3, the slope (which is related to $\phi_{0}$ according to Eq. \ref{Iphfar}) of the curves is the same within experimental error. Variations in $\phi_{0}$ can therefore be ruled out. Fluctuations of $A_{dark}$ can also be excluded using the following reasoning: a change in $A_{dark}$ will induce a change in $d$ via the feedback loop, and thus a change in the \textit{shape} of $I_{dark}(V)$\cite{footnote7}. This change is not clearly observed (see Fig. \ref{figure6}a). The only remaining possible sources of current fluctuations are changes in $A_{ph}$, and to some extent $\phi _{m}$. Since $A_{dark}$ (and thus $S$) are constant, changes in $A_{ph}$ must originate from changes of the prefactor $K_{ph}$ of Eq. \ref{Iphfar}.

The same analysis, performed for all distances, shows that: i) at large distances (point a of Fig. \ref{figure5}) the instabilities are relatively small. This may be because the local electric field is smaller and thus (photo)electro-chemical modification of the surface oxide layer is less important, ii) for reduced distances before mechanical contact, the instabilities mostly concern $K_{ph}$, iii) for measurements performed under mechanical contact (point e of Fig. \ref{figure5}) the instabilities are very large. This is again consistent with the notion that greater electric fields result in an increase in oxide modifying (photo)electro-chemical processes. These findings are confirmed by the quantitative analysis presented in section V.B, where a possible microscopic interpretation will be proposed. 

\begin{figure}[tbp]
\includegraphics[clip,width=7.5 cm] {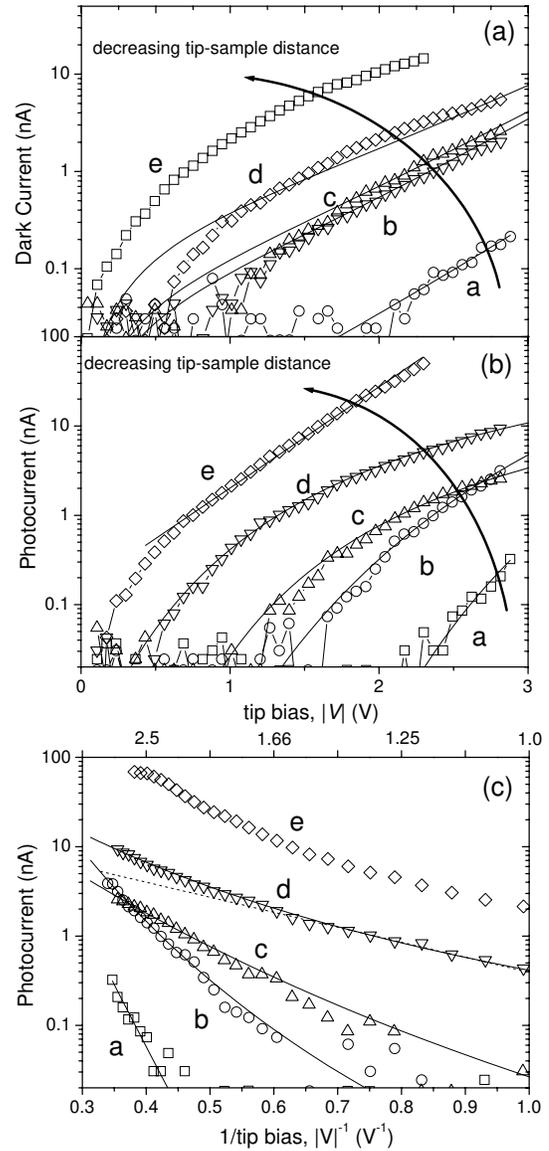}
\caption{Plot of the dark current (a) and of the photocurrent (b) as a function of $V$ for the sharp tip shown in Fig. \ref{figure6}. (c) Photocurrent spectra versus $1/V$ show exponential behaviour over almost two orders or magnitude. This, and the increase in slope with increasing tip/sample distance (from curve d to curve a), indicate a FN-like tunneling process for photoelectrons. The excess photocurrent observed at high tip bias on curve d is also consistent with this interpretation. Solid black curves are fits to the data using Eq. \ref{Idark} for the dark current and Eq. \ref{Iphmid} for the photocurrent (see section V).}
\label{figure8}
\end{figure}

\subsection{Tunneling processes}

The bias dependence of the dark current for different distances is shown in Fig. \ref{figure8}a, while that of the tunneling photocurrent is shown in Figs. \ref{figure8}b and \ref{figure8}c. For clarity, only curves for which the photocurrent is largest are presented, although the results are typical of a much larger number of experiments. It is apparent that the tunneling dark current and the tunneling photocurrent originate from distinct mechanisms. The tunneling dark current tends to present a flatter spectral response than the photocurrent when plotted against tip bias on a log-linear scale. Furthermore, the slope of the tunneling dark current is relatively insensitive to a change in distance whereas the slope of the tunneling photocurrent spectra increases greatly with an increase in distance. 

For curves a, b, and c in Fig. \ref{figure8}c, the dependence of $\ln(I_{ph})$ as a function of $1/V$ is quasi-linear over two orders of magnitude, consistent with Eq. \ref{Iphfar} and/or Eq. \ref{Iphmid} in the small bias regime. The tunneling photocurrent behaviour is thus consistent with the FN-like process described in section II whereas it will be seen that the tunneling dark current is better described by a standard tunneling process from the semiconductor valence band. 

Qualitatively, the behavior of the threshold bias for the onset of the FN-like photocurrent, $V_{th}$, as a function of distance can only be understood by including image charge effects. Experimentally, $V_{th}$ is estimated at -2.6 V, -1.6 V, and -1.15 V for curves a, b, and c in Fig. \ref{figure8}b respectively. Eq. \ref{seuilloin} which neglects image charge effects, predicts $V_{th}\approx -4.3$V (using $\phi_{m}$ = 5.4 eV for a clean gold surface) which is too large and independent of bias. On the other hand, the experimentally observed thresholds can be understood by using Eq. \ref{seuil} with tip/sample distances of 1.56 nm, 1.0 nm and 0.9 nm for curves a, b and c respectively.

For curve d of Fig. \ref{figure8}c, $\ln(I_{ph})$ above 1.5 V is larger than the linear extrapolation from lower bias (dotted line). Despite being in the near contact regime (see point d of Fig. \ref{figure5}), the excess current is not caused by electrostatic forces acting on the cantilever, as seen from the absence of a bias-dependent atomic force. Rather, it is the result of a lowering of the tunnel barrier due to image charge effects according to Eq. \ref{Iphmid}.

In the case of curve e of Fig. \ref{figure8}b and Fig. \ref{figure8}c, the photocurrent is large. It is estimated that several $10^{-3}$ of the total number of photoelectrons created in the tip are injected into the metal. No linear dependence of $\ln(I_{ph})$ as a function of $1/V$ is observed over the whole bias range. In Fig. \ref{figure8}b, curve e above a tip bias of 0.5 V shows a simple exponential dependence on tip bias, consistent with the vanishing of the tunnel barrier\cite{chen}, and will be discussed further below. Since the tip is now in mechanical contact with the surface (see Fig. \ref{figure5}), the width of the tunneling gap is constant and equal to the oxide thickness and only the area of the surface indentation, $S$, changes with set point. When in mechanical contact, it is indeed observed experimentally that $I_{ph}(V)$ and $I_{dark}(V)$ change only by a multiplicative factor with a change in set point\cite{footnote10}. 

The dark current bias dependence is found to be approximately exponential at large tip/sample distances (see curves a, b and c of Fig. \ref{figure8}a). This dependence, which occurs over a small bias range, can be understood by expanding Eq. \ref{d0dark} to first order in $V$.

\section{Interpretation}

\subsection{Quantitative analysis of the $I_{ph}\left(V\right)$ and $I_{dark}\left( V\right)$ bias dependences}

A simultaneous fitting procedure, with equal weight, of $I_{ph}(V)$ and of $I_{dark}(V)$ was used that consisted of selecting initial values of $d$, $\phi_{0}$, $\phi _{m}$, $A_{dark}$ and $A_{ph}$, and in finding the set of values for these quantities which minimizes $R=\sum\limits_{i}\left[I_{ph}\left(V_{i}\right) - I_{ph}^{calc}\left(V_{i}\right)\right]^{2} +
\sum\limits_{i}\left[I_{dark}\left( V_{i}\right) -I_{dark}^{calc}\left(V_{i}\right)\right]^{2}$ where the index $i$ labels the individual data points. The calculated values $I_{ph}^{calc}\left(V_{i}\right)$ and $I_{dark}^{calc}\left(V_{i}\right)$ were obtained using Eq. \ref{Iphstandard} and Eq. \ref{Iphmid} for the photocurrent, and Eq. \ref{Idark} for the dark current (for which the integral was evaluated numerically). Although there are five fitting parameters, the fit is severely constrained by the fact that both currents depend on $d$, $\phi _{0}$, and $\phi _{m}$ explicitely, as well as indirectly on $A_{dark}$ via the feedback process. This co-dependence allows for an independent determination of $d$ and $\phi_{0}$ which is not possible in many works on field emission\cite{bonard02}. The fits, shown as solid lines in Fig. \ref{figure8}, are in satisfactory agreement with both the experimental results and the qualitative analysis of the preceding section. The excess photocurrent in curve d at large bias is also accounted for. For the dark current, the small but systematic discrepancy near the threshold may be attributed to the surface barrier $V_{b}$ at the tip apex. If $|V| < V_{b}$, the tunneling dark current is reduced because majority carriers must also tunnel across the space charge layer, which is estimated\cite{footnote11} to have a width of order 20 nm. 

The values of the parameters obtained using the above procedure, and corresponding to the various curves of Fig. \ref{figure6}c and Fig. \ref{figure8} are summarized in Table 1.

\subsection{Origin of current instabilities}
Examination of the results of Table 1 concerning curve c of Fig. \ref{figure6}c confirms the qualitative analysis of section IV.A and suggests the following comments on the nature of the current instabilities:

i) $d$ takes relatively stable values around 1.2 nm. The magnitude of $\lambda $ is of the order of 0.4 eV. According to Eq. \ref{seuil} the lowering of the barrier due to image charge effects cannot be neglected. 

ii) The values obtained for $\phi_{0}$ and $\phi_{m}$ are reasonable: $\phi_{0}$ lies between 3.7 eV and 4.2 eV and $\phi _{m}$ lies between 3.3 eV and 5.2 eV. 

iii) The largest instability concerns the quantity $A_{ph}$, which fluctuates by a factor of 30 between the various curves. $\phi_{m}$ also fluctuates by about 1.5 eV. In contrast, the fluctuations of $A_{dark}$, $\phi_{0}$ and $d$ are significantly smaller. Since $A_{dark}$ (and thus $S$) does not fluctuate, variations in $A_{ph}$ are related to changes in $K_{ph}$. 

The instabilities are presumably related to the fact that the experiments take place in a slightly humid, gaseous environment. With a clean tip and sample surface in vacuum the only possible mechanism for instabilities is the desorption of gold or silicon atoms under the effect of the electric current. This process may explain the observed changes in $\phi_{m}$ if desorbed silicon atoms are adsorbed onto the gold surface, but cannot explain the instabilities of $K_{ph}$ which themselves are most reasonably associated with the presence of an oxide layer covering the tip apex. The thickness ($d_{ox}$) of the oxide is not known, but it is assumed to be smaller than the smallest value found for $d$ in Table \ref{table1}, $\approx$ 1 nm (the thickness\cite{zhang} of the planar natural oxide ranges between 0.5 and 1.5 nm). 

\begin{table}[tbp]
\caption{Fitting parameters for the curve fits shown in Fig. \ref{figure8}, and for six spectra measured at point c in Fig. \ref{figure6}. $A_{ph}$ influences only the photocurrent, while $d$, $\phi_{0}$, and $\phi_{m}$ are shared between the two according to Eq. \ref{Iphstandard}, Eq. \ref{Iphmid} and Eq. \ref{Idark}. $A_{dark}$ is an indirectly shared parameter via the feedback loop.}
\label{table1}
\begin{ruledtabular}
\begin{tabular}{lccccc}
spectrum & $A_{dark}$ (A) & $A_{ph}$ (A) & $d$ (nm) & $\phi_{0}$ & $\phi_{m}$ \\
\hline
a & 656 & $1.9\times 10^{-6}$ & 1.52 & 4.4 & 4.5\\
\hline
b& 171 & $4.1\times 10^{-6}$ & 1.29 & 5.1 & 6.7\\
\hline
c(i)& 4.4 & $6.5\times 10^{-9}$ & 1.27 & 4.2 & 5.3\\
c(ii)& 5.4 & $7.8\times 10^{-10}$ & 1.18 & 3.8 & 3.3\\
c(iii)& 5.4 & $5.6\times 10^{-9}$ & 1.23 & 4.1 & 4.7\\
c(iv)& 5.1 & $3.6\times 10^{-9}$ & 1.25 & 4.1 & 4.9\\
c(v)& 3.9 & $2\times 10^{-10}$ & 1.17 & 3.7 & 3.2\\
c(vi)& 4.5 & $4.5\times 10^{-9}$ & 1.21 & 4.2 & 4.9\\
\hline
d& 1 & $1.85\times 10^{-8}$ & 1.05 & 4.6 & 5.7\\
\end{tabular}
\end{ruledtabular}
\end{table}

\begin{figure}[tbp]
\includegraphics[clip,width=7 cm] {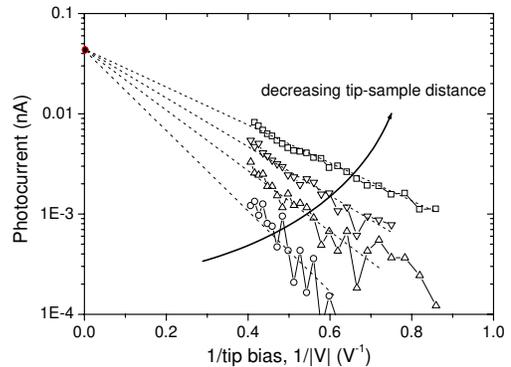}
\caption{Plot of the photocurrent as a function of $1/V$ for the blunt tip shown in Fig. \ref{figure7}. The shared origin indicates that $A_{ph}$ is nearly identical for all four curves taken at different tip/sample distances.}
\label{figure10}
\end{figure}

Since $\epsilon_{SiO_{2}}$ = 4 the electric field is mainly dropped in the tunneling gap between the oxide surface and the metal. The relevant distance for dark current tunneling is still the \textit{total} distance, $d$, between the silicon surface of the tip and the metal, whereas for FN-like tunneling which is an electric field driven effect, it is $d - d_{ox}$. In the simplest case where image charge effects are neglected, this results only in a modification of $A_{ph}$:
\begin{equation}
A_{ph}^{\ast}=A_{ph}\exp(\kappa_{ph}d_{ox}),
\label{Aphstar}
\end{equation}
where $\kappa_{ph}$ is defined in Eq. \ref{Iphfar}. Eq. \ref{Aphstar} provides a possible explanation for the observed variations in $A_{ph}$ as any small variation in $d_{ox}$ translates into large fluctuations in the photocurrent prefactor, $A_{ph}^{\ast}$. In the case of the blunt tip, variations in $A_{ph}$ are averaged out over a large surface area: when $I_{ph}$ is plotted against $1/V$ for different tip/sample distances as shown in Fig. \ref{figure10}, the extrapolated curves pass through the same point at $1/V = 0$ indicating that $A_{ph}$ is nearly identical for all distances. It is partly in consequence of this that fluctuations in the photocurrent (see Fig. \ref{figure7}b) are minimal.

The presence of an oxide layer and the inclusion of image charge effects may alter the average $\overline{\phi}$ over the oxide and vacuum barriers and therefore $\kappa_{ph}$, but a detailed analysis is unreasonable given the additional adjustable parameters (e.g. oxide thickness, dielectric constant and effective mass) which are not known if the oxide is ultra-thin. The impact of an oxide layer on the results was evaluated using a perturbation-like approach which consisted of introducing an extremely thin oxide (0.1 to 0.2 nm) into the equations of gradually increasing thickness, and repeating the numerical fit described above. While $A_{dark}$ was increased significantly, the most physically meaningful parameters ($d$, $\phi_{0}$) were unchanged so the main conclusions of the present work are unaffected if, as assumed above, $d_{ox} \lesssim$ 0.5 nm.

Investigations of instabilities in field emission\cite{miyamoto03} suggest a possible microscopic mechanism for fluctuations in $d_{ox}$: the tip is thought to be covered by a layer of adsorbed molecules or ions (water, oxygen, etc...) that diffuse into (and out of) the oxide layer under the effect of an electric field, thereby changing its effective thickness. In view of the fact that tunneling occurs via only a handful of atoms/molecules at the tip apex, the observed sharp changes in tunneling current between certain well-defined values can be attributed to the adsorption or desorption of individual molecules. Other explanations involving charge build-up at local defects in the oxide\cite{rogers85, farmer87} can be discounted since this would imply a change in effective barrier height which, in the majority of cases, is not observed.

\subsection{Effect of tip/sample distance}

From Table \ref{table1} $d$ decreases as expected from curve a to curve d. The change in slope of the curves is greater than an $\exp(-d/V)$ dependence and arises from the dependence of $\lambda$ and thus $\kappa _{ph}$ on distance. 

$A_{dark}$ decreases steadily by almost 3 orders of magnitude when $d$ is decreased. This might be attributed to a progressive reduction in $S$ with decreasing distance, but a similar trend in $A_{ph}$ which also depends linearly on $S$, is not clearly observed perhaps due to the fluctuations in $A_{ph}$. It may be that $A_{dark}$ is under-estimated at small $d$ due to the presence of the oxide layer which is neglected in the model (see the preceding sub-section). At small $d$ the relative effect of the oxide is larger and will result in an artificial reduction of $A_{dark}$. Further studies are necessary to clarify whether this is the case, or whether the trend in $A_{dark}$ is due to a change in $K_{dark}$ (i.e. the orbital overlaps) and/or $S$. 

In the case of curve e of Fig. \ref{figure8}b, taken under mechanical tip/sample contact, the exponential bias dependence of the tunneling photocurrent can be explained assuming a complete vanishing of the tunnel barrier. Assuming that $d$ is smaller than half the de Broglie wavelength of the electron, and taking a square barrier, it has been recognized (see Fig. 2.6 of Ref. \onlinecite{chen}) that the probablility of ballistic transmission is an exponential function of the barrier height. Taking the barrier height to be the spatial average calculated using Eq. \ref{phibar}, the probability for ballistic transmission is of the form
\begin{equation} T=\exp \left\{-(\overline{\phi}-E_{g})/\phi^{\ast}\right\},
\label{trans}
\end{equation}
where $\phi^{\ast}$ is some characteristic energy. Again using Simmons' approximation for the image charge potential, Eq. \ref{phiparab}, the ballistic photocurrent is of the form \begin{equation}
I_{ph}=A_{ph}\exp \left\{-(\chi_{0}/2 +2.8\lambda -aqV^{\ast}/2)/\phi
^{\ast}\right\}. \label{Iphnear}
\end{equation} 
The barrier given by Eq. \ref{phiparab} only vanishes if $\lambda$ is larger than about 0.5 eV (this assumes for simplicity that $\phi_{0} = \phi_{m}$ = 4 eV). Taking an oxide thickness of 0.5 nm as suggested by section V.B, and $\epsilon_{SiO_{2}}$ = 4, Eq. \ref{lambda} yields $\lambda=0.25$ eV which is too small. However the dielectric constant of the ultra-thin oxide layer may be smaller than the bulk value, so $\lambda\approx 0.5$ eV is reasonable. Using this value and taking the oxide effective mass as given in Ref. \onlinecite{brar96}, a de Broglie wavelength for $V$ = -2 V of the order of 3 nm is obtained, which is indeed larger than twice $d_{ox}$. Similarly, using Fig. 2.6 of Ref. \onlinecite{chen}, $\phi^{\ast}$ is calculated to be 0.025 eV. In view of uncertainties in the value of $d_{ox}$ and in the shape of the barrier, the agreement with the experimental value of 0.1 eV can be considered as satisfactory.

\begin{figure}[tbp]
\includegraphics[clip,width=7 cm] {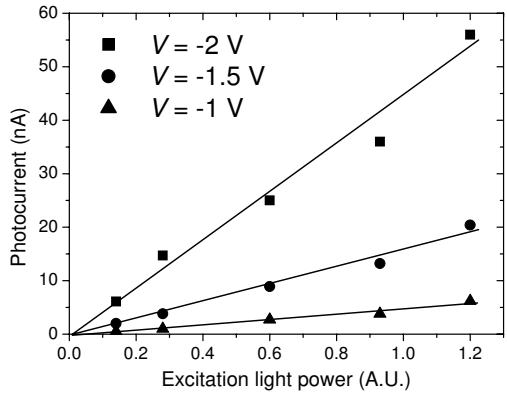}
\caption{Dependence of the photocurrent on excitation light power. For the three values of the tip bias, this dependence is linear.}
\label{figure9}
\end{figure}

\subsection{Comparison with other works}

In the Nijmegen model, all tunneling occurs via midgap surface states uniformly distributed in energy and is formulated for low tip bias\cite{prins96}. This mechanism, illustrated in Fig. \ref{figure2} (arrow 3), can be excluded here for two reasons: i) the most significant tip bias dependence of the tunneling current in the Nijmegen model arises via the bias dependence of the surface depletion layer width\cite{gartner59}, and is too weak to explain the strong variation observed in this work. Extension of the Nijmegen model to include the bias dependence of $\kappa _{ph}$, given by Eq. \ref{d0photo1}, is also unable to explain the results, and ii) the tunneling photocurrent in the Nijmegen model depends \textit{logarithmically} on the incident light power in contrast with the linear dependence observed here (see Fig. \ref{figure9}).

Besides the tip bias values used in Nijmegen work, which are too small to observe FN-like tunneling, their use of GaAs tips is also significant. In defect free GaAs the surface recombination velocity is orders of magnitude larger than that for defect free silicon\cite{smith78}. Consequently, the characteristic time for trapping at midgap states may be faster than the characteristic tunneling time from the conduction band so tunneling will proceed via midgap states. The opposite may be true in silicon so that FN-like tunneling is favored.

\section{Conclusion}

Photo-assisted tunneling between a p-type silicon tip illuminated from the rear and a metallic gold surface has been studied for absolute tip bias up to 3 V. The additional tunneling current induced by light excitation, due to injection of photocarriers from the tip into the surface, is large and comparable with the dark current. Its magnitude depends \textit{linearly} on the light excitation power and no surface photovoltage is observed.

The tunneling dark and photocurrent spectra are distinct and are quantitatively interpreted by a simple model that yields reasonable values of the tip and metal work functions and of the tip/sample distance. The tunneling dark current is described by a standard process between the semiconductor valence band and the metallic density of states. In contrast, the tunneling photocurrent behavior is explained by a electric-field-dependent FN-like process between the semiconductor conduction band and the metallic density of states, including image charge effects. This mechanism accounts for all the results obtained before mechanical tip/sample contact : the $\exp(1/V)$ dependence of the tunnel photocurrent at large tip/sample distance, as well as the excess current at large tip bias observed just before contact. Once in contact the tunnel barrier for photoelectrons vanishes, and current flow is determined by the probablility of ballistic emission over the barrier. This results in a photocurrent dependent exponentially on $V$.

For spin injection applications with GaAs tips\cite{pierce88,divincenzo99,datta90}, it is possible that at low bias, injection will be best described by the standard-tunneling-based Nijmegen model. However, the present work indicates that FN-like tunneling may also be observed at higher tip bias, even with GaAs. FN-like tunneling is likely to be spin dependent since the vacuum states into which photoelectrons tunnel should be hybridized with the spin polarized states of the magnetic surface. In this case, FN-like tunneling would then be the relevant process (for example) for studies of the highly-polarized minority spin $d$-states of Fe and Co surfaces which lie about 2 eV above the Fermi level\cite{stroscio95}.

In common with many works on field emission from silicon tips and on transport through Si/SiO$_{2}$ structures, current instabilities are observed. However, since the instabilities tend to occur over timescales longer than that taken for the spectral measurements of both the tunneling dark current and the tunneling photocurrent, a meaningful analysis of the spectra is possible. With the aid of the model developed here, it is possible to show that the current instabilities are mostly related to changes in the thickness of the oxide layer covering the tip apex. Changes in this thickness may be related to electric field induced adsorption or desorption of foreign species such as water or oxygen. For this reason, an appropriate surface passivation of the tip is essential for future spin-polarized injection investigations using GaAs tips in liquid or neutral gas environments. Studies are underway to optimize such a tip treatement\cite{berkovits06}.

\acknowledgements{The authors thank P. Prod'homme for the provision of electrochemically grown gold-on-silicon samples, F. Da Costa for help in the building of the experimental setup, and J. Peretti for helpful discussions.}

\bibliographystyle{apsrev}
\bibliography{bibrowe}

\begin{thebibliography}{48}
\bibitem[{foo({\natexlab{a}})}]{footnote1}
\bibinfo{note}{The geometry in which the tip is excited from the planar rear
  surface of the tip is favorable for the elimination of the parasitic optical
  effects that have plagued all efforts aimed at observing spin-dependent
  tunnel processes between a semiconducting tip and a magnetic surface.}
\bibitem[{foo({\natexlab{b}})}]{footnote2}
\bibinfo{note}{This approximation is straightforward for photoelectrons for
  which the kinetic part of the energy is very small, so that $E \approx
  E_{z}$. For the dark tunneling current, the corrected form of Eq. \ref{Idark}
  is obtained by replacing $E_{z}$ by $E\cos^{2}\theta $ and $d$ by
  $d$/$\cos\theta \ $where $\theta $ is the angle between \ the tip axis and
  the electron momentum $k$. For realistic values of $d$ and $\phi _{0}$, the
  tunneling current rapidly decreases with $\theta $, with a half-width at half
  maximum $\theta_{m}\approx 15^{\circ}$. Since $\theta _{m}$ is
  weakly-dependent both on $E$ and $V^{\ast}$, and since $cos\theta _{m}$ is
  close to unity, the only effect of the angular dependence of the tunneling
  probability is to add a bias-independent prefactor, proportional to
  $\theta_{m}$.}
\bibitem[{foo({\natexlab{c}})}]{footnote3}
\bibinfo{note}{As seen from Eq. \ref{phibar}, the value of $\overline{\phi}$
  depends on the spatial limits $d_{1}$ and $d_{2}$ of the barrier.
  Simmons\cite {simmons63} takes $d_{1}$ and $d_{2}$ at the Fermi level. Here,
  expressions for $d_{1}$ and $d_{2}$ are calculated at the energy $E_{g}$ of
  the tunneling photoelectrons, and are found by replacing $\phi _{0}$ by
  $\phi_{0}-E_{g}$ in Eq. 36 of the Ref \onlinecite{simmons63}.}
\bibitem[{foo({\natexlab{d}})}]{footnote13}
\bibinfo{note}{The tip bias should be large enough that $d_{2}$ be determined
  by Eq. 38 of Ref. \onlinecite{simmons63}, otherwise Eq. 37 of the same work
  should be applied. The condition, $qV > \phi_{0}$, cited by Simmons for the
  crossover between the two regimes, corresponding to the threshold bias for
  FN-like tunneling, is erroneous. By equating the expressions given in Eq. 37
  and Eq. 38, and replacing $\phi_{0}$ by the smaller $\chi_{0}$, it is found
  that Eq. 38 of Ref. \onlinecite{simmons63} is valid even at tip bias as low
  as 1 V, especially when image charge effects (i.e. $\lambda$) are large. A
  rather inelegant expression for $V_{th}$ is obtained.}
\bibitem[{foo({\natexlab{e}})}]{footnote14}
\bibinfo{note}{Using Eq. 37 of Ref. \onlinecite{simmons63}, the modification of
  the spatial average of the tunneling barrier due to image charge effects is
  found to be negligible.}
\bibitem[{foo({\natexlab{f}})}]{footnote15}
\bibinfo{note}{Since at this energy the light absorption length is comparable
  with the tip height, some electrons are created in the sub-apex region.
  However, because of the high tip doping level, only a small fraction of the
  photocurrent originates from electrons photocreated in the depletion zone and
  the photocurrent can still be assumed to be independent of bias.}
\bibitem[{foo({\natexlab{g}})}]{footnote7}
\bibinfo{note}{The overall amplitude of $I_{dark}(V)$ is imposed by the
  feedback loop, and cannot be considered as an adjustable parameter.}
\bibitem[{foo({\natexlab{h}})}]{footnote10}
\bibinfo{note}{Since the $I_{ph}(V)$ and $I_{dark}(V)$ spectra change only by a
  multiplicative factor, the potential drop in the tip and cantilever is still
  negligible with respect to the applied tip bias.}
\bibitem[{foo({\natexlab{i}})}]{footnote11}
\bibinfo{note}{The depletion layer width at the tip apex is estimated
  numerically at almost a factor of 10 larger than that calculated for a planar
  contact.}
\end{thebibliography}

\appendix

\section{Negligible field enhancement at the tip apex}

When discussing field emission phenomena, it is usual to describe the effect of tip geometry on the electric field by introducing the enhancement factor $\gamma$ defined by $F_{apex}=\gamma \frac{V}{d}$, where $F_{apex}$ is the electric field at the tip apex. There is currently great interest in the use of sharp tips and high aspect ratio objects such as carbon nanotubes, where field enhancement factors up to $\gamma = 10^4$ are reported\cite{bonard02}, since these make for very efficient field emitters. In the present case $\gamma$ is of order unity despite the fact that sharp tips are used. The main difference here is that the tip/sample distance $d$ is of order 1 nm (i.e. smaller than the tip radius, $r\approx 60$ nm), whereas in field emission studies the cathode/anode distance can be as large as several $\mu$m (i.e. much larger than the tip radius). At the scale of 1 nm therefore, the ``sharp'' tips used in this study are \textit{locally flat} and $\gamma \approx$ 1.

\begin{figure}
\includegraphics[clip,width=8 cm] {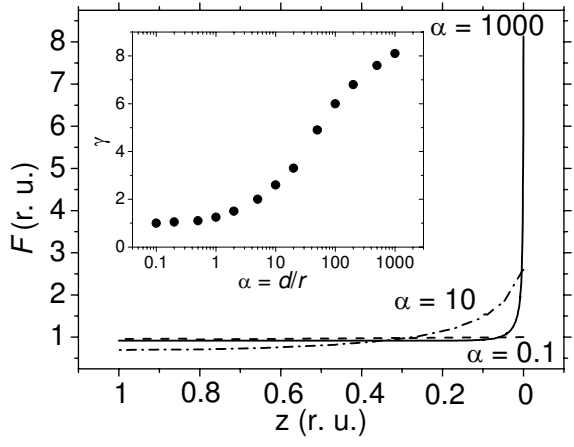}
\caption{Variation in normalized electric field (units of $V$/$d$) along the axis from the
tip (z = 0) to the sample surface (z = 1 reduced unit). A large increase in electric field is observed close to the tip for large $\protect\alpha$. The inset shows the variation in $\gamma$ with the aspect ratio, $\alpha = d/r$ . For small values of $\alpha$, corresponding to the experiments reported here, $\gamma \rightarrow$ 1.}
\label{efieldtip}
\end{figure}

The electric field F(z) has been calculated along the axis between the tip apex and the surface. This calculation was performed numerically in two dimensions, using a finite element resolution of Laplace's equation, with a fixed potential difference $V$ between tip and sample. The tip is modelled as a conical section of half angle $12^\circ$, similar to the value for the tips used in the experiments, and the tip apex is terminated by a circle of radius $r$. As shown in Fig. \ref{efieldtip}, for $\alpha = 10^{3}$, which corresponds to a typical field emission experiment, the electric field is  greatly enhanced near the tip surface where $\gamma \approx 8$. For $\alpha = 10^{-1}$ which corresponds to the present tunneling experiments, F(z) is constant and $\gamma = 1$. The variation in $\gamma$ with $\alpha$ is shown in Fig. \ref{efieldtip}(inset). It is noted that $\gamma$ only deviates from unity for $\alpha$ larger than 1.

\end{document}